\documentclass{article}%
\usepackage[top=3cm, bottom=3cm, left=3cm, right=3cm]{geometry}
\usepackage{amsmath}
\usepackage{amsfonts}
\usepackage{amssymb}
\usepackage{graphicx}%
\providecommand{\U}[1]{\protect\rule{.1in}{.1in}}
\begin{document}

\title{Can the $\psi(4040)$ explain the peak associated with $Y(4008)$?}
\author{Milena Piotrowska$^{1}$, Francesco Giacosa$^{1,2}$ and Peter Kovacs$^{3}$\\\textit{$^{1}$Institute of Physics, Jan Kochanowski University, }\\\textit{ul. Swietokrzyska 15, 25-406, Kielce, Poland.}\\\textit{$^{2}$Institute for Theoretical Physics, J. W. Goethe University, }\\\textit{ Max-von-Laue-Str. 1, 60438 Frankfurt, Germany.}\\\textit{$^{3}$Institute for Particle and Nuclear Physics,},\\\textit{Wigner Research Centre for Physics,}\\\textit{Hungarian Academy of Sciences, H-1525 Budapest, Hungary }}
\date{}
\maketitle

\begin{abstract}
We study the well-known resonance $\psi(4040)$,
corresponding to a $3^{3}S_{1}$ charm-anticharm vector state $\psi(3S)$,
within a QFT approach, in which the decay channels into
$DD$, $D^{\ast}D$, $D^{\ast}D^{\ast}$, $D_{s}D_{s}$ and $D_{s}^{\ast}D_{s}$
are considered. The spectral function shows sizable
deviations from a Breit-Wigner shape (an enhancement, mostly
generated by $DD^{\ast}$ loops, occurs); moreover, besides the $c\bar{c}$ pole of $\psi(4040)$, a second dynamically generated broad pole at $4$ GeV
emerges. Naively, it is tempting to identify this new pole with
the unconfirmed state $Y(4008).$ Yet, this state was not
seen in the reaction $e^{+}e^{-}\rightarrow\psi(4040)\rightarrow DD^{\ast}$,
but in processes with $\pi^{+}\pi^{-}J/\psi$ in the final state. A
detailed study shows a related but different mechanism: a broad peak at
$4$ GeV in the process $e^{+}e^{-}\rightarrow\psi(4040)\rightarrow DD^{\ast
}\rightarrow\pi^{+}\pi^{-}J/\psi$ appears when $DD^{\ast}$ loops are
considered. Its existence in this reaction is not
necessarily connected to the existence of a dynamically generated pole, but
the underlying mechanism - the strong coupling of $c\bar{c}$ to
\ $DD^{\ast}$ loops - can generate both of them. Thus, the
controversial state $Y(4008)$ may not be a genuine resonance, but a peak
generated by the $\psi(4040)$ and $D^{\ast}D$ loops
with $\pi^{+}\pi^{-}J/\psi$ in the final state.

\end{abstract}

\section{Introduction}

The understanding of the nature and properties of hadronic states is a
substantial challenge for both experimentalists and theorists. Remarkable
progress in the field of charmonium spectroscopy was provided in the past
decades: while various resonances emerge as conventional charmonium states
(ordinary $\bar{c}c$ objects), the so-called $X,$ $Y,$ and $Z$ states are
candidates for exotic hadrons (such as molecules, hybrids, multi-quarks
objects or glueballs; see Refs. \cite{rev,rev2016,pillonireview,nielsen} and refs. therein).

In this work, we shall concentrate on the vector sector in the energy region
close to $4$ GeV. Here, the well established charmonium vector state
$\psi(4040)$ is listed in the Particle Data Group (PDG) \cite{pdg2018} (it has
$J^{PC}=1^{--}$ where, as usual, $P$ refers to parity and $C$ to charge
conjugation). This resonance can be successfully interpreted as a charmonium
state with $(n,L,S,J)=(3,0,1,1)$, where $n$ is the principal number, $L$ the
angular momentum, $S$ the spin and $J$ the total spin); hence, the
nonrelativistic spectroscopic notation reads $n$ $^{2S+1}L_{J}=$ $3$~
$^{3}S_{1}$ (see e.g. Refs. \cite{isgur, eichten, segoviarev, segovia2} and
refs. therein).

Very close to $4$ GeV, the enigmatic (and not yet confirmed) resonance
$Y(4008)$ was also observed as a significant enhancement by the Belle
Collaboration when measuring the cross section of $e^{+}e^{-}\rightarrow
\pi^{+}\pi^{-}J/\psi$ via initial state radiation (ISR) technique \cite{belle}
and later on confirmed by the same group \cite{belle2}: its mass was
determined as $4008\pm40_{-28}^{+114}$ MeV and the decay width as
$\Gamma=226\pm44\pm87$ MeV. Moreover, a broad $Y(4008)$ was also found in\ the
recent analysis of Ref.  \cite{gao}. However, the statistic at Belle was
pretty limited and $Y(4008)$ could \textit{not} be confirmed by subsequent
experiments studying the same production process at BaBar \cite{babar} and
BESIII \cite{besiii}, making its existence rather controversial. Nevertheless,
several possible theoretical assignments on its nature have been suggested,
including non-conventional scenarios as $D^{\ast}\bar{D}^{\ast}$ molecular
state \cite{molecular, molecular2} (see however also Ref. \cite{ding}),
tetraquark \cite{tetraquark1, tetraquark2} or even an interference effect with
background \cite{interference}. Moreover, in Refs.
\cite{charmstate1,charmstate2} it was proposed to identify $Y(4008)$ as
$\psi(3S)$ charmonium state, but this assignment is not favoured, since, as
mentioned above, $\psi(4040)$ is well described by a standard $\psi(3S)$
state. The unexplained status of the observed structure corresponding to
$Y(4008)$ makes it an interesting subject that deserves clarification, hence
we aim to perform a detailed study of the nearby energy region.

To this end, we develop a quantum field theoretical effective model in which a
\textit{single} $c\bar{c}$ seed state, to be identified with $\psi(4040),$
couples to $DD,$ \ $DD^{\ast},$ and $D^{\ast}D^{\ast}.$ The immediate question
is if we can describe both resonances $\psi(4040)$ and $Y(4008)$ at the same
time and within a unique setup. The idea that we test is somewhat reminiscent
of studies in the light scalar sector, in which the state $a_{0}(980)$ can be
seen as a companion pole of the predominantly $q\bar{q}$ state $a_{0}(1450)$
\cite{dullemond,tornqvist,boglione,a0} as well as the light $\kappa$ state, now
named $K_{0}^{\ast}(700),$ as a companion pole of the $K_{0}^{\ast}(1430)$
\cite{milenathomas}. Quite interestingly, two poles emerge also in the study
of the charmonium resonance $\psi(3770)$ \cite{psi3770}. As we shall see, some
similarities, but also some important differences, will emerge between those
studies and the one that we are going to present.

As a first step of our analysis, we calculate the spectral function of
$\psi(4040)$. As expected, it is peaked at about $4.04$ GeV, but it cannot be
approximated by a standard Breit-Wigner shape; most remarkably, it may develop
an additional enhancement below $4$ GeV (this is due to the strong coupling of
the bare $\bar{c}c$ state to the $DD^{\ast}$ channel). Moreover, two poles
appear in the complex plane: one corresponding to the peak in the spectral
function (hence to $\psi(4040)$) and an additional companion pole, dynamically
generated by meson-meson (mostly $DD^{\ast}$) quantum fluctuations. A
large-$N_{c}$ study shows that $\psi(4040)$ behaves as a conventional
quark-antiquark state while the enhancement does not fit into this standard picture.

At a first sight, it appears quite natural to assign the state $Y(4008)$ to
this additional dynamically generated pole. Yet, a closer inspection is
necessary: the study of the decay chain in which $Y(4008)$ was seen,
$e^+e^- \rightarrow \psi(4040)\rightarrow DD^{\ast}\rightarrow\pi^{+}\pi^{-}J/\psi$ (the latter
can occur via a light scalar state, most notably $f_{0}(500)$, but not only), which shows that a broad peak at about $3.9$ GeV emerges for the cross-section
$e^{+}e^{-}\rightarrow\pi^{+}\pi^{-}J/\psi$ (also when $e^{+}e^{-}$ comes from
a previous ISR process, as observed in experiment). This is due to the fact
that the loop contribution of $DD^{\ast}$ is peaked at about $m_{D}%
+m_{D^{\ast}}\simeq3.9$ GeV. As we shall explain in detail later on, this
contribution is multiplied by the modulus squared of the propagator of
$\psi(4040)$, centered at $4.04$ GeV and about $80$ MeV large, hence a sizable
overlap is present. As we shall show, the emergent peak at about $3.9$ GeV is
very far from a Breit-Wigner state, but is rather distorted. Strictly
speaking, the very existence of an additional companion pole is not necessary
for the emergence of this signal, but both phenomena arise from a strong
coupling of the seed state to $DD^{\ast}$, hence it is rather natural that
they both take place at the same time.

The paper is organized as follows: In Sec. 2 we introduce theoretical model,
in particular the Lagrangians, the possible decays channels of $\psi(4040)$
resonance with corresponding theoretical expression for decay widths, loop
function (hence, the propagator) and spectral function. Moreover, we show in
detail the determination of the parameters of our model. In Sec. 3 we present
our results. Summary and outlooks are presented in Sec. 4. Additional results
for different parameter values are reported in the Appendices.

\section{The model}

In this section we present the theoretical model used to analyze the energy
region close to $4$ GeV. In our approach, only a \textit{single} standard
$c\bar{c}$ seed state corresponding to $\psi(4040)$ is included.

\subsection{Theoretical framework}

The resonance $\psi(4040)$ can be described by a relativistic interaction
Lagrangian that couples it to its decay products [two pseudoscalar mesons
($DD$ and $D_sD_s$), one vector and one pseudoscalar meson ($DD^{\ast}$ and $D^*_sD_s$), and two vector
mesons ($D^{\ast}D^{\ast}$)]:%

\begin{equation}
\mathcal{L}_{\psi(4040)}=\mathcal{L}_{VPP}+\mathcal{L}_{VPV}+\mathcal{L}%
_{VVV}\label{lag}%
\end{equation}
where
\begin{equation}
\mathcal{L}_{VPP}=ig_{\psi DD}\psi_{\mu}\left[  \left(  \partial^{\mu}%
D^{+}\right)  D^{-}+\left(  \partial^{\mu}D^{0}\right)  \bar{D}^{0}+\left(
\partial^{\mu}D_{s}^{+}\right)  D_{s}^{-}\right]  +h.c.\text{ },\label{lag1}%
\end{equation}%
\begin{equation}
\mathcal{L}_{VPV}=ig_{\psi D^{\ast}D}\tilde{\psi}_{\mu\nu}\left[
\partial^{\mu}D^{\ast+\nu}D^{-}+\partial^{\mu}D^{\ast0\nu}\bar{D}^{0}%
+\partial^{\mu}D_{s}^{\ast+\nu}D_{s}^{-}\right]  +h.c.\text{ },\label{lag2}%
\end{equation}%
\begin{equation}
\mathcal{L}_{VVV}=ig_{\psi D^{\ast}D^{\ast}}\left[  \psi_{\mu\nu}\left(
D^{\ast+\mu}D^{\ast-\nu}+D^{\ast0\mu}\bar{D}^{\ast0\nu}+D_{s}^{\ast+\mu}%
D_{s}^{\ast-\nu}\right)  \right]  +h.c.\text{ }.\label{lag3}%
\end{equation}
The quantities $g_{\psi DD}$, $g_{\psi D^{\ast}D}$, $g_{\psi D^{\ast}D^{\ast}%
}$ are the coupling constants that are determined by using experimental data
from PDG \cite{pdg2018}, see Sec. 2.2 for details. Moreover $\psi_{\mu\nu
}=\partial_{\mu}\psi_{\nu}-\partial_{\nu}\psi_{\mu}$ and $\tilde{\psi}_{\mu
\nu}=\frac{1}{2}\varepsilon_{\mu\nu\rho\sigma}\psi^{\rho\sigma}$ are the
vector-field tensor and its dual. In particular, the term $\mathcal{L}_{VPP}$
describes the decay processes $\psi(4040)\rightarrow D^{+}D^{-}$,
$\psi(4040)\rightarrow D^{0}\bar{D}^{0}$ and $\psi(4040)\rightarrow D_{s}%
^{+}D_{s}^{-}$, the term $\mathcal{L}_{VPV}$ the processes $\psi
(4040)\rightarrow D^{\ast0}\bar{D}^{0}+h.c.$, $\psi(4040)\rightarrow D^{\ast
+}D^{-}+h.c.$ and $\psi(4040)\rightarrow D_{s}^{\ast+}D_{s}^{-}+h.c.$, and,
finally, the term $\mathcal{L}_{VVV}$ the transitions $\psi(4040)\rightarrow
D^{\ast+}D^{\ast-}$ and $\psi(4040)\rightarrow D^{\ast0}\bar{D}^{\ast0}$. The masses of the particles are taken from the PDG: $m_{D^+}=m_{D^-}=1869.65 \pm 0.05$ MeV, $m_{D^0}=m_{\bar{D}^0}=1864.83 \pm 0.05$ MeV, $m_{D^{*0}}=m_{\bar{D}^{*0}}=2006.85 \pm 0.05$ MeV, $m_{D^{*+}}=m_{D^{*-}}=2010.26 \pm 0.05$ MeV, $m_{D^+_s}=m_{D^-_s}=1968.34 \pm 0.07$ MeV and $m_{D_s^{*+}}=m_{D_s^{*-}}=2112.2 \pm 0.4$ MeV. Other decay channels (as for instance $D^*_sD^*_s$) are not considered because they are kinematically forbidden.

As usual, the theoretical expressions for the tree-level decay widths for each
type of decay can be obtained from the Feynman rules and read (by keeping the
mass of the decaying state as `running' and denoted by $m$)
\begin{equation}
\Gamma_{\psi\rightarrow D^{+}D^{-}+h.c}(m)=\frac{\left[  k(m,m_{D^{+}%
},m_{D^{-}})\right]  ^{3}}{6\pi m^{2}}g_{\psi DD}^{2}F_{\Lambda}%
(k)\hspace{0.1cm},
\end{equation}%
\begin{equation}
\Gamma_{\psi\rightarrow D^{\ast+}D^{-}+h.c}(m)=\frac{2}{3}\frac{\left[
k(m,m_{D^{\ast+}},m_{D^{-}})\right]  ^{3}}{\pi}g_{\psi D^{\ast}D}%
^{2}F_{\Lambda}(k)\hspace{0.1cm},
\end{equation}%
\begin{equation}
\Gamma_{\psi\rightarrow D^{\ast+}D^{\ast-}}(m)=\frac{2}{3}\frac{\left[
k(m,m_{D^{\ast+}},m_{D^{\ast-}})\right]  ^{3}}{\pi m_{D^{\ast+}}^{2}}g_{\psi
D^{\ast}D^{\ast}}^{2}\left[  2+\frac{\left[  k(m,m_{D^{\ast+}},m_{D^{\ast-}%
})\right]  ^{2}}{m_{D^{\ast+}}^{2}}\right]  F_{\Lambda}(k)\hspace{0.1cm}.
\end{equation}
The quantity
\begin{equation}
k\equiv|\vec{k}|\equiv k(m,m_{A},m_{B})=\frac{\sqrt{\left(  m^{2}-m_{A}%
^{2}-m_{B}^{2}\right)  ^{2}-4m_{A}^{2}m_{B}^{2}}}{2m}\label{momentum}%
\end{equation}
is the modulus of the three-momentum of one of the outgoing mesons $A$ or $B,$
with masses $m_{A}$ and $m_{B}$ respectively, in the rest frame of the
decaying particle with mass $m$. The tree-level on-shell decay width for the
state $\psi(4040)$ are obtained by setting $m=m_{\psi(4040)}=4.04$ GeV (here
and in the following, we use the average mass $4039.6\pm4.3$ MeV
\cite{pdg2018} rounded to $4.040\pm0.004$ GeV).

Another important quantity is the vertex function (or form factor)
$F_{\Lambda}(k)$, which assures that each quantity calculated in our model is finite. Note, one could include the vertex function directly in the Lagrangian
by making it nonlocal \cite{nonlocal} (covariance can be also preserved
\cite{nonlocal2}). In our study we employed a Gaussian form factor
\begin{equation}
F_{\Lambda}\equiv F_{\Lambda}^{\text{Gauss}}(k)=e^{-2\frac{k^{2}}{\Lambda^{2}%
}}\text{ ,}\label{gaussian}%
\end{equation}
which emerges in microscopic approaches such as $^{3}P_{0}$ mechanisms (which models the creation of quark-antiquark pairs in the QCD vacuum) used in
quark models \cite{3p0old,3p0new}. However, there are other possibilities of
choosing the cutoff function, the basic requirements being a \textit{smooth}
behavior (a step function is not an admissible choice) and a
\textit{sufficiently fast} decrease on the real positive axis. For
completeness, another smooth form factor
\begin{equation}
F_{\Lambda}\equiv F_{\Lambda}^{\text{Dipolar}}(k)=\left(  1+\frac{k^{4}%
}{\Lambda^{4}}\right)  ^{-2}\label{dipolar}%
\end{equation}
has been tested here in order to check how the results depend on the choice of
this function. As we shall see, there are no substantial changes.

What is rather important is the numerical value of $\Lambda$. We expect a
value between $\sim0.4$ and $\sim0.8$ GeV. Namely, for the light $\kappa$ meson,
$\Lambda\simeq0.5$ GeV was obtained by a fit to data \cite{milenathomas}. In
the recent work of Ref. \cite{psi3770}, a even smaller value $\Lambda
\approx0.3$ GeV is found (but a value of about $0.4$ GeV also delivers results
compatible with data).\ A comparison with the $^{3}P_{0}$ model induces a
cutoff of $\Lambda\approx0.8$ GeV \cite{3p0old,3p0new} (but that approach
was typically not employed to calculate meson-meson loops). In this work, we
test how the results vary upon changing $\Lambda$ in the range from $0.4$ to
$0.8$ GeV (and for different form factors), but only up to $0.6$ GeV
physically acceptable results are obtained. 

It should be stressed that our approach is an effective model of QCD,
therefore the value of $\Lambda$ does \textit{not} represent the maximal value
for the possible values of the momentum $k.$ When $k$ is larger than $\Lambda
$, then that particular decay is suppressed, but this is a physical
consequence of the nonlocal interaction between the decaying meson and its
decay products (all of them being extended objects). The momentum $k$ can take
any value from $0$ to $\infty$, even arbitrarily larger than $\Lambda.$ In
particular, the normalization of the spectral function (a crucial feature of
our approach, see below) involves an integration up to $k\rightarrow\infty$.
Of course, even if it is allowed to take $k$ arbitrarily large, this does not
imply that the model is physically complete: since we take into account a
single resonance, the $\psi(4040)$, our approach can describe some of the
features around $4$ GeV (and up to about $4.15$ GeV). Above that, one should
include the resonance $\psi(4160)$ and, even further, the resonance
$\psi(4415).$ (For completeness, we have tested the case in which $\psi(4040)$
and $\psi(4160)$ are present at the same time. As we shall comment later on,
including $\psi(4160)$ does not substantially change the results for
$\psi(4040)$).

Next, the scalar part of the propagator of the vector field $\psi_{\mu}$ is
\begin{equation}
\Delta_{\psi}(p^{2}=m^{2})=\frac{1}{m^{2}-M_{\psi}^{2}+\Pi(m^{2}%
)+i\varepsilon}\text{ ,}\label{prop}%
\end{equation}
where $M_{\psi}$ is the bare mass of the vector state $\psi(4040)$ (to be
identified with the seed $\bar{c}c$ mass in absence of loop corrections). The
quantity $\Pi(m^{2})=\operatorname{Re}(\Pi(m^{2}))+i\operatorname{Im}%
(\Pi(m^{2}))$ is the one-particle irreducible self-energy. At the one-loop
level, $\Pi(m^{2})$ is the sum of all one-loop contributions:
\begin{align}
\Pi(m^{2}) &  =\Pi_{D^{+}D^{-}}(m^{2})+\Pi_{D^{0}\bar{D}^{0}}(m^{2}%
)+\Pi_{D_{s}^{+}D_{s}^{-}}(m^{2})+\Pi_{D^{\ast0}\bar{D}^{0}+h.c}%
(m^{2})\nonumber\\
&  +\Pi_{D^{\ast+}D^{-}+h.c}(m^{2})+\Pi_{D_{s}^{\ast+}D_{s}^{-}+h.c}%
(m^{2})+\Pi_{D^{\ast0}\bar{D}^{\ast0}}(m^{2})+\Pi_{D^{\ast+}D^{\ast-}}%
(m^{2})+...,
\end{align}
where dots refer to further subleading contributions of other small decay
channels. Moreover, the imaginary part $\operatorname{Im}(\Pi(m^{2}))$ reads
(optical theorem)
\begin{align}
\operatorname{Im}(\Pi(m^{2})) &  =m\Big(\Gamma_{\psi(4040)\rightarrow
DD}(m)+\Gamma_{\psi(4040)\rightarrow D_{s}D_{s}}+\Gamma_{\psi(4040)\rightarrow
D^{\ast}D}(m)\nonumber\\
&  +\Gamma_{\psi(4040)\rightarrow D_{s}^{\ast}D_{s}}(m)+\Gamma_{\psi
(4040)\rightarrow D^{\ast}D^{\ast}}(m)\Big)\text{ ,}%
\end{align}
where:
\begin{equation}
\Gamma_{\psi(4040)\rightarrow DD}(m)=\Gamma_{\psi(4040)\rightarrow D^{+}D^{-}%
}(m)+\Gamma_{\psi(4040)\rightarrow D^{0}\bar{D}^{0}}(m)\text{ ,}\label{dd}%
\end{equation}%
\begin{equation}
\Gamma_{\psi(4040)\rightarrow D^{\ast}D}(m)=\Gamma_{\psi(4040)\rightarrow
D^{\ast+}D^{-}+h.c}(m)+\Gamma_{\psi(4040)\rightarrow D^{\ast0}\bar{D}^{0}%
+h.c}(m)\text{ ,}\label{dsd}%
\end{equation}%
\begin{equation}
\Gamma_{\psi(4040)\rightarrow D^{\ast}D^{\ast}}=\Gamma_{\psi(4040)\rightarrow
D^{\ast+}D^{\ast-}}+\Gamma_{\psi(4040)\rightarrow D^{\ast0}\bar{D}^{\ast0}%
}\text{ .}\label{dsds}%
\end{equation}

The real part $\operatorname{Re}(\Pi(m^{2}))$ is calculated by dispersion
relations. For instance, for the decay channel $\psi(4040)\rightarrow
D^{+}D^{-}$ one has:
\begin{equation}
\operatorname{Re}(\Pi_{D^{+}D^{-}}(m^{2}))=-\frac{1}{\pi} PP \int
\limits_{2m_{D^{+}}}^{\infty}2m^{\prime}\frac{m^{\prime}\Gamma_{\psi
(4040)\rightarrow D^{+}D^{-}}(m^{\prime})}{m^{2}-m^{\prime2}}\mathrm{dm}%
^{\prime}\text{ ;}%
\end{equation}
(similar expressions hold for all other channels). \begin{figure}[h]
\begin{center}
\includegraphics[width=0.52 \textwidth]{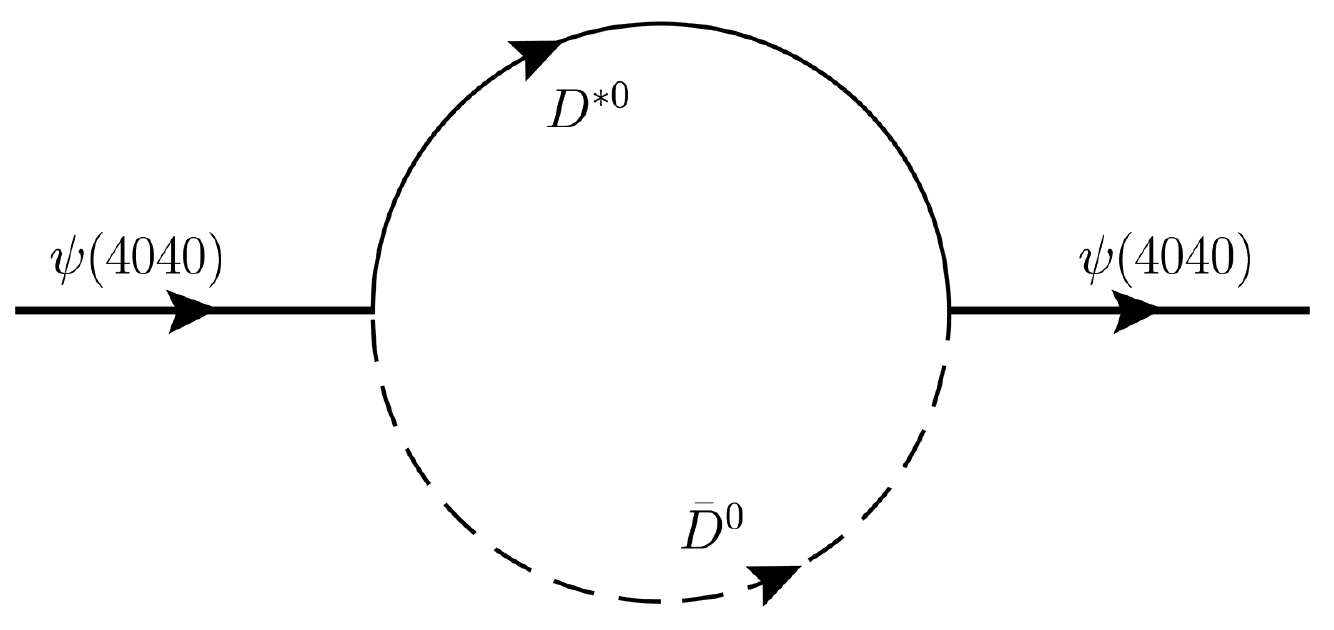}
\end{center}
\caption{Example of one-loop contribution. Here the case of $D^{\ast0}$ and
$\bar{D}^{0}$ is shown.}%
\end{figure}The spectral function is connected to the imaginary part of the
propagator introduced above as
\begin{equation}
d_{\psi}(m)=\frac{2m}{\pi}\left\vert \operatorname{Im}\Delta_{\psi}%
(p^{2}=m^{2})\right\vert \text{ .}\label{spectralfunct}%
\end{equation}
The quantity $d_{\psi}(m)\mathrm{dm}$ determines the probability that the
state $\psi(4040)$ has a mass between $m$ and $m+\mathrm{dm}$. It must fulfill
the normalization condition
\begin{equation}
\int\limits_{0}^{\infty}d_{\psi}(m)\mathrm{dm}=1\text{ .}\label{norm}%
\end{equation}
\qquad The validity of the normalization is a crucial feature of our study,
since it guarantees unitarity \cite{lupo}. It is a consequence of our
theoretical approach (for a detailed mathematical proof of its validity, see
Ref. \cite{scalarboson}). Note, in Eq. (\ref{norm}) the integration is up to
$m\rightarrow\infty$ (hence, $k\rightarrow\infty,$ see Eq. (\ref{momentum})).
In practice, we shall verify numerically that Eq. (\ref{norm}) is fulfilled
(we do so by integrating up to $10$ GeV, far above the region of interest of
about 4 GeV).

In addition, the partial spectral functions read:
\begin{align}
d_{\psi\rightarrow DD}(m)  &  =\frac{2m}{\pi}\left\vert \Delta_{\psi}%
(m^{2})\right\vert ^{2}m\Gamma_{\psi(4040)\rightarrow DD}(m)\text{ ,}\\
d_{\psi\rightarrow D_{s}^{+}D_{s}^{-}}(m)  &  =\frac{2m}{\pi}\left\vert
\Delta_{\psi}(m^{2})\right\vert ^{2}m\Gamma_{\psi(4040)\rightarrow D_{s}D_{s}%
}(m)\text{ ,}\\
d_{\psi\rightarrow DD^{\ast}}(m)  &  =\frac{2m}{\pi}\left\vert \Delta_{\psi
}(m^{2})\right\vert ^{2}m\Gamma_{\psi(4040)\rightarrow DD^{\ast}}(m)\text{
,}\\
d_{\psi\rightarrow D_{s}^{\ast}D_{s}}(m)  &  =\frac{2m}{\pi}\left\vert
\Delta_{\psi}(m^{2})\right\vert ^{2}m\Gamma_{\psi(4040)\rightarrow D_{s}%
^{\ast}D_{s}}(m)\text{ ,}\\
d_{\psi\rightarrow D^{\ast}D^{\ast}}(m)  &  =\frac{2m}{\pi}\left\vert
\Delta_{\psi}(m^{2})\right\vert ^{2}m\Gamma_{\psi(4040)\rightarrow D^{\ast
}D^{\ast}}(m)\text{ .}%
\end{align}
For instance, $d_{\psi\rightarrow DD^{\ast}}(m)\mathrm{dm}$ is the probability
that the resonance $\psi(4040)$ has a mass between $m$ and $m+\mathrm{dm}$ and
decays in the channel $DD^{\ast}$ \cite{duecan}. Similar interpretations hold
for all other channels. These quantities are physically interesting since they
emerge when different channels are studied; if, for instance, the process
$e^{+}e^{-}\rightarrow DD^{\ast}$ is considered, the corresponding cross
section is proportional to $d_{\psi\rightarrow DD^{\ast}}(m)$.

\subsection{Determination of the parameters}

Our model contains five free parameters: the three coupling constants $g_{\psi
DD}$, $g_{\psi D^{\ast}D}$, $g_{\psi D^{\ast}D^{\ast}}$ entering Eqs.
(\ref{lag1}), (\ref{lag2}), and (\ref{lag3}), the bare mass of the vector
state $M_{\psi}$ entering in the propagator (\ref{prop}), and the energy scale
(cutoff) $\Lambda$ included in Eq. (\ref{gaussian}) or Eq. (\ref{dipolar}).

We proceed as follows: first, we fix the value of $\Lambda$ in the range
$0.4$-$0.6$ GeV. Then, in order to determine the coupling constants three
experimental values are needed. We use the following measured ratios of
branching fractions reported in PDG \cite{pdg2018} (see also Refs.
\cite{babar,wang,belle3,cleo}):
\begin{equation}
\left.  \frac{\mathcal{B}(\psi(4040)\rightarrow D\bar{D})}{\mathcal{B}%
(\psi(4040)\rightarrow D^{\ast}\bar{D})}\right\vert _{\exp}=0.24\pm
0.05\pm0.12\text{ ,}%
\end{equation}%
\begin{equation}
\left.  \frac{\mathcal{B}(\psi(4040)\rightarrow D^{\ast}\bar{D}^{\ast}%
)}{\mathcal{B}(\psi(4040)\rightarrow D^{\ast}\bar{D})}\right\vert _{\exp
}=0.18\pm0.14\pm0.03\text{ ,}%
\end{equation}
where the first error is statistical and the second is systematic. Moreover, we
employ the experimental value of the total width of the $\psi(4040)$ resonance
PDG \cite{pdg2018}
\begin{equation}
\Gamma_{\psi(4040)}^{\text{tot,exp}}=80\pm10\hspace{0.2cm}\text{MeV .}%
\end{equation}
The vector state $\psi(4040)$ decays into various two-body final states. The
decay channels contributing mostly to its total decay width are: $DD$,
$D_{s}D_{s}$, $D^{\ast}D$, $D_{s}^{\ast}D_{s}$ and $D^{\ast}D^{\ast}$. The
corresponding theoretical expression for the total decay width of $\psi(4040)$
state is given by
\begin{equation}
\Gamma_{\psi(4040)}^{\text{tot,theory}}=\Gamma_{\psi(4040)\rightarrow
DD}^{\text{on}\hspace{0.1cm}\text{shell}}+\Gamma_{\psi(4040)\rightarrow
D_{s}D_{s}}^{\text{on}\hspace{0.1cm}\text{shell}}+\Gamma_{\psi
(4040)\rightarrow D^{\ast}D}^{\text{on}\hspace{0.1cm}\text{shell}}%
+\Gamma_{\psi(4040)\rightarrow D_{s}^{\ast}D_{s}}^{\text{on}\hspace
{0.1cm}\text{shell}}+\Gamma_{\psi(4040)\rightarrow D^{\ast}D^{\ast}%
}^{\text{on}\hspace{0.1cm}\text{shell}}\text{ ,}%
\end{equation}
where \textquotedblleft on-shell\textquotedblright\ means that the physical
PDG mass $m=4.04$ GeV is employed.

Finally, the coupling constants $g_{\psi DD}$, $g_{\psi D^{\ast}D}$ and
$g_{\psi D^{\ast}D^{\ast}}$ as well as their errors are obtained upon
minimizing the $\chi^2$ function $F_{E}$ depending on all this three parameters:
\begin{align}
F_{E}(g_{\psi DD},g_{\psi D^{\ast}D},g_{\psi D^{\ast}D^{\ast}})  &  =\left(
\frac{\frac{\Gamma_{\psi\rightarrow DD}^{theory}(g_{\psi DD})}{\Gamma
_{\psi\rightarrow D^{\ast}D}^{theory}(g_{\psi D^{\ast}D})}-\left.
\frac{\mathcal{B}(\psi(4040)\rightarrow D\bar{D})}{\mathcal{B}(\psi
(4040)\rightarrow D^{\ast}\bar{D})}\right\vert _{\exp}}{\delta\left.
\frac{\mathcal{B}(\psi(4040)\rightarrow D\bar{D})}{\mathcal{B}(\psi
(4040)\rightarrow D^{\ast}\bar{D})}\right\vert _{\exp}}\right)  ^{2}%
+\nonumber\\
\left(  \frac{\frac{\Gamma_{\psi\rightarrow D^{\ast}D^{\ast}}^{theory}(g_{\psi
D^{\ast}D^{\ast}})}{\Gamma_{\psi\rightarrow D^{\ast}D}^{theory}(g_{\psi
D^{\ast}D})}-\left.  \frac{\mathcal{B}(\psi(4040)\rightarrow D^{\ast}\bar
{D}^{\ast})}{\mathcal{B}(\psi(4040)\rightarrow D^{\ast}\bar{D})}\right\vert
_{\exp}}{\delta\left.  \frac{\mathcal{B}(\psi(4040)\rightarrow D^{\ast}\bar
{D}^{\ast})}{\mathcal{B}(\psi(4040)\rightarrow D^{\ast}\bar{D})}\right\vert
_{\exp}}\right)  ^{2}  &  +\left(  \frac{\Gamma_{\psi(4040)}^{tot,theory}%
(g_{\psi DD},g_{\psi D^{\ast}D},g_{\psi D^{\ast}D^{\ast}})-\Gamma_{\psi
(4040)}^{tot,exp}}{\delta\Gamma_{\psi(4040)}^{tot,exp}}\right)  ^{2}%
\end{align}
The bare mass $M_{\psi}$ was fixed under the requirement that the maximum of
the spectral function corresponds to the nominal mass of $\psi(4040)$, hence
to $4.04$ GeV.

The numerical values of $g_{\psi DD}$, $g_{\psi D^{\ast}D}$, $g_{\psi D^{\ast
}D^{\ast}}$ and of bare mass $M_{\psi}$ are reported in Tab. 1 in Sec. 3.1 for
given values of the cutoff $\Lambda$. As expected, $g_{\psi D^{\ast}D}$,
$g_{\psi D^{\ast}D^{\ast}}$ depend rather mildly on $\Lambda$, but $g_{\psi
DD}$ quite strongly. However, the decay width into $DD$ is very small and affects
only slightly the overall picture. For completeness, we report in Appendix A
also the partial decay widths for various choices of the cutoff and for
different form factors. While the results are basically compatible with each
other, future experimental determination of the channel $\psi(4040)\rightarrow
D_{s}^{+}D_{s}^{-}$ would be very helpful to constrain our model.

\section{Results}

In this section we show the results and comment on them. First, in Sec. 3.1 we
concentrate on the form of the full spectral function (as well as the partial
ones into $DD$, $D^{\ast}D$, $D^{\ast}D^{\ast}$ channels) of the resonance
$\psi(4040)$. Moreover we determine the position of the pole(s) in the complex
plane. Then, in Sec. 3.2 we present the discussion of the important process
$e^{+}e^{-}\rightarrow J/\psi\pi^{+}\pi^{-}$ and the possible generation of a
peak for $Y(4008)$.

\subsection{Spectral function and pole positions}

Since scattering data have still quite large errors, it is not yet possible to
determine the value of $\Lambda$ through a fit. Moreover, such a fit would
also need to include an unknown background contribution. This is why $\Lambda$
has been varied in a quite large range in\ Table 1, in which the positions of
the poles have also been reported. As already mentioned in the introduction, we always find two poles in the complex plane, one corresponding to the maximum of the spectral function and an additional dynamically generated one. For $\Lambda$ up to $\sim0.5$ GeV similar
results are obtained, but for larger values the second pole (even if always
present) appears at higher values. We have also tested values larger than
$0.6$ GeV, but they do not generate satisfactory results. (This outcome is in
agreement with the results of Sec. \ref{subsec32} and Appendix B, see later on). 

In the following, we choose for the numerical value $\Lambda=0.42,$ since it
generates a pole whose imaginary part is $40$ MeV, then $\Gamma_{\psi
(4040)}^{\text{pole }}=80$ MeV.\ We then use this value for illustration and
for the presentation of the plots. (Yet, it should not be considered as a
sharp value for the cutoff). The spectral function $d_{\psi}(m)$ defined
in\ Eq. (\ref{spectralfunct}) is shown in\ Fig. 2.a together with a standard
Breit-Wigner function peaked at $4.04$ GeV and with a width of $80$ MeV, which
serves for comparison.

\begin{figure}[h] 
\begin{center}
\begin{minipage}[b]{7.70cm}
\centering
\includegraphics[width=7.55cm]{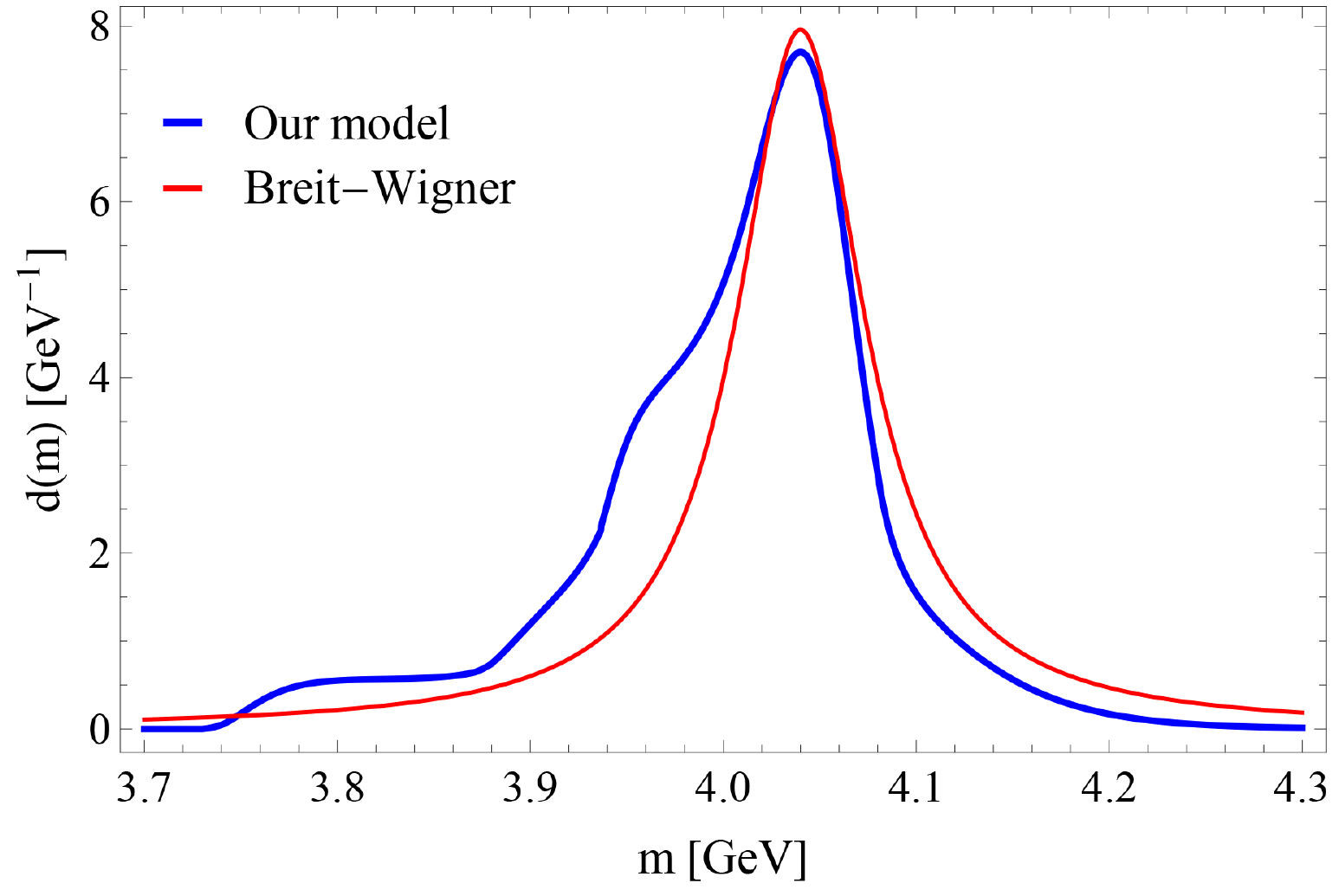}\\\textit{a)}
\end{minipage}
\begin{minipage}[b]{7.70cm}
\centering
\includegraphics[width=7.70cm]{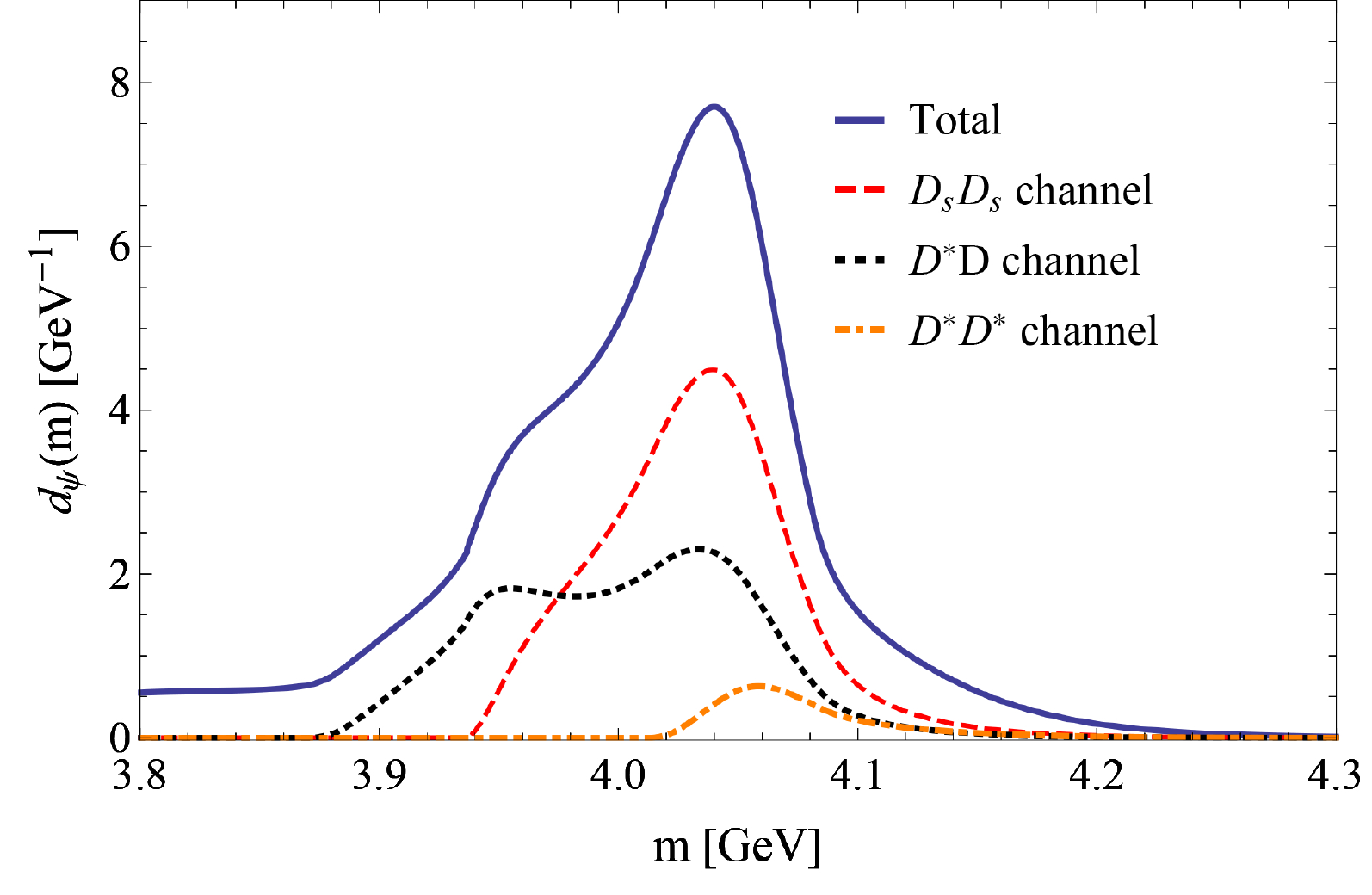}\\\textit{b)}
\end{minipage}
\end{center}
\caption{Panel (a) presents the shape of the spectral function of the
resonance $\psi(4040)$ of Eq. (\ref{spectralfunct}) (blue line) with
comparison to the standard Breit-Wigner form (red line). Panel (b) presents
the partial spectral functions for $D_{s}D_{s}$, $D^{\ast}D$ and $D^{\ast
}D^{\ast}$ channels (This channels are described by Eq. (\ref{dd}), Eq.
(\ref{dsd}) and Eq. (\ref{dsds})). The corresponding parameters are presented
in Table 1.}%
\label{rys1}%
\end{figure}Only one single peak close to $4.04$ GeV corresponding to the
standard seed $\bar{c}c$ state is present. While the Breit-Wigner function
approximates quite well the spectral function close to the peak, sizable
deviations close to $3.9$ GeV are present. This is due to an enhancement in
the energy region below $4$ GeV, which is generated by meson-meson loops.

In\ Fig. 2.b we present the contributions of individual channels ($DD$,
$D^{\ast}D$ and $D^{\ast}D^{\ast}$) to the total spectral function. The
$D^{\ast}D$ channel turns out to be the most important for the deformation on
the l.h.s. of the spectral function. In the complex plane we found two poles:
one for $4.053-0.039$ $i$ GeV, corresponding to $\psi(4040)$ resonance, and
one for $3.934-0.030$ $i$ GeV.\ Thus, even if only one single seed state
identified with $\psi(4040)$ was included into the calculations, two poles
naturally emerge.

At a first sight, it is  tempting to identify the additional pole with the
controversial state $Y(4008).$ Moreover, a look at Table 1 shows that a second
additional pole always exists, and up to values of about $0.5$ GeV the
dynamically generated pole is not far from $4$ GeV. However, care is needed
for the following reasons: the pole width of the additional state is too small
when compared to the experimental value (about $200$ MeV), but it should be
stressed that a direct comparison of this pole width and the experiment is
misleading, since very different reactions were measured, see later on. If the
enhancement in\ Fig. 1 is real, it should be visible in the cross-section of
the channel $e^{+}e^{-}\rightarrow\psi(4040)\rightarrow D^{\ast}D,$ which is
proportional to $d_{\psi\rightarrow DD^{\ast}}(m)$ (see Fig. 1.b); presently,
the data have too large errors to resolve such a complicated structure. Quite
importantly, the state $Y(4008)$ has been observed in the ISR reaction
$e^{+}e^{-}\rightarrow\gamma e^{+}e^{-}\rightarrow\gamma\pi^{+}\pi^{-}J/\psi$
and not in the $DD^{\ast}$ channel, see next section for the discussion of
this important point.

\begin{table}[h]
\centering
\par
\renewcommand{\arraystretch}{1.25}
\begin{tabular}
[c]{|c|c|c|c|c|}\cline{2-5}%
\multicolumn{1}{c|}{} & \multicolumn{2}{c|}{\textbf{Gaussian form factor}} &
\multicolumn{2}{c|}{\textbf{Dipolar form factor}}\\\hline
$\Lambda$ & Parameters & Pole(s) [GeV] & Parameters & Pole(s) [GeV]\\\hline
0.4 & $g_{\psi DD}=48.8 \pm4.6$ & $\bullet\hspace{0.3cm} (4.052 \pm0.003)$ &
$g_{\psi DD}=25.4 \pm5.0$ & $\bullet\hspace{0.3cm} (4.058 \pm0.019)$\\
& $g_{\psi D^{*}D}=3.60 \pm0.95$ & $\hspace{0.3cm} -i (0.035 \pm0.005)$ &
$g_{\psi D^{*}D}=3.50 \pm0.58$ & $\hspace{0.3cm} -i(0.050 \pm0.014)$\\
& $g_{\psi D^{*}D^{*}}=1.65 \pm0.86$ & $\blacklozenge\hspace{0.3cm} (3.936
\pm0.005)$ & $g_{\psi D^{*}D^{*}}=1.93 \pm0.89$ & $\blacklozenge\hspace{0.3cm}
(3.941 \pm0.003)$\\
& $M_{\psi}=4.00$ & $\hspace{0.3cm} -i(0.022 \pm0.001)$ & $M_{\psi}=4.02$ &
$\hspace{0.3cm} -i(0.045 \pm0.010)$\\\hline
0.42 & $g_{\psi DD}=39.8 \pm5.0$ & $\bullet\hspace{0.3cm} (4.053 \pm0.004)$ &
$g_{\psi DD}=21.7 \pm4.4$ & $\bullet\hspace{0.3cm} (4.062 \pm0.023)$\\
& $g_{\psi D^{*}D}=3.44 \pm0.80$ & $\hspace{0.3cm} -i(0.039 \pm0.009)$ &
$g_{\psi D^{*}D}=3.06 \pm0.49$ & $\hspace{0.3cm} -i(0.056 \pm0.011)$\\
& $g_{\psi D^{*}D^{*}}=1.85 \pm0.93$ & $\blacklozenge\hspace{0.3cm}(3.934
\pm0.006)$ & $g_{\psi D^{*}D^{*}}=1.94 \pm0.89$ & $\blacklozenge\hspace{0.3cm}
(3.942 \pm0.004)$\\
& $M_{\psi}=4.01$ & $\hspace{0.3cm} -i(0.030 \pm0.001)$ & $M_{\psi}=4.03$ &
$\hspace{0.3cm} -i(0.052 \pm0.010)$\\\hline
0.45 & $g_{\psi DD}=29.9 \pm5.0$ & $\bullet\hspace{0.3cm} (4.055 \pm0.005)$ &
$g_{\psi DD}=17.4 \pm3.8$ & $\bullet\hspace{0.3cm}(4.070 \pm0.027) $\\
& $g_{\psi D^{*}D}=3.14 \pm0.61$ & $\hspace{0.3cm} -i(0.047 \pm0.018)$ &
$g_{\psi D^{*}D}=2.57 \pm0.38$ & $\hspace{0.3cm} -i(0.066 \pm0.008)$\\
& $g_{\psi D^{*}D^{*}}=2.07 \pm0.99$ & $\blacklozenge\hspace{0.3cm} (3.928
\pm0.008)$ & $g_{\psi D^{*}D^{*}}=1.97 \pm0.89$ & $\blacklozenge\hspace{0.3cm}
(3.943 \pm0.006)$\\
& $M_{\psi}=4.02$ & $\hspace{0.3cm} -i(0.042 \pm0.002)$ & $M_{\psi}=4.04$ &
$\hspace{0.3cm} -i(0.064 \pm0.011)$\\\hline
0.5 & $g_{\psi DD}=19.5 \pm4.2$ & $\bullet\hspace{0.3cm} (4.055 \pm0.009)$ &
$g_{\psi DD}=12.6 \pm3.0$ & $\bullet\hspace{0.3cm} (4.087 \pm0.033)$\\
& $g_{\psi D^{*}D}=2.64 \pm0.39$ & $\hspace{0.3cm} -i(0.066 \pm0.054)$ &
$g_{\psi D^{*}D}=2.02\pm0.27 $ & $\hspace{0.3cm} -i(0.083 \pm0.006)$\\
& $g_{\psi D^{*}D^{*}}=2.3 \pm1.0 $ & $\blacklozenge\hspace{0.3cm} (3.918
\pm0.007)$ & $g_{\psi D^{*}D^{*}}=2.02 \pm0.89$ & $\blacklozenge\hspace{0.3cm}
(3.943 \pm0.011)$\\
& $M_{\psi}=4.04$ & $\hspace{0.3cm}-i(0.063 \pm0.004)$ & $M_{\psi}=4.05$ &
$\hspace{0.3cm} -i(0.085 \pm0.014)$\\\hline
0.6 & $g_{\psi DD}=10.4 \pm2.7$ & $\bullet\hspace{0.3cm} (4.025 \pm0.015)$ &
$g_{\psi DD}=7.4 \pm2.0$ & $\bullet\hspace{0.3cm} (4.032 \pm0.019)$\\
& $g_{\psi D^{*}D}=1.95 \pm0.22$ & $\hspace{0.3cm} -i(0.041 \pm0.031)$ &
$g_{\psi D^{*}D}=1.44 \pm0.16$ & $\hspace{0.3cm} -i(0.035 \pm0.020)$\\
& $g_{\psi D^{*}D^{*}}=2.3 \pm1.0$ & $\blacklozenge\hspace{0.3cm} (4.056
\pm0.017)$ & $g_{\psi D^{*}D^{*}}=2.09 \pm0.90$ & $\blacklozenge\hspace{0.3cm}
(4.056 \pm0.023)$\\
& $M_{\psi}=4.07$ & $\hspace{0.3cm} -i(0.032 \pm0.007)$ & $M_{\psi}=4.08$ &
$\hspace{0.3cm} -i(0.029 \pm0.006)$\\\hline
\end{tabular}
\caption{Position of the poles in the complex plane for different parameters
used in the model. The coupling constants $g_{\psi DD}$ and $g_{\psi D^{\ast
}D^{\ast}}$ are dimensionless, $g_{\psi D^{\ast}D}$ has dimensions GeV$^{-1}$,
$\Lambda$ and $M_{\psi}$ are in GeV.}%
\label{tablepar}%
\end{table}

Next, we perform a large-$N_{c}$ study of the resonance $\psi(4040)$ (where
$N_{c}$ refers to the number of colors in QCD). To this end, we introduce the
scaling parameter $\lambda\in(0,1)$, linked to $N_{c}$ as $\lambda=3/N_{c}$,
and consider the scaling of the coupling constants as \cite{largencwitten}
\begin{equation}
g_{\psi DD}\rightarrow\sqrt{\lambda}g_{\psi DD}\text{ , }g_{\psi D^{\ast}%
D}\rightarrow\sqrt{\lambda}g_{\psi D^{\ast}D}\text{ , }g_{\psi D^{\ast}%
D^{\ast}}\rightarrow\sqrt{\lambda}g_{\psi D^{\ast}D^{\ast}}\text{ }.
\end{equation}
Clearly, by setting $\lambda=1$, we reobtain our physical results. In the
opposite limiting case, $\lambda=0$, the spectral function reduced to a delta
function centered in the seed mass, $\delta(m-M_{\psi})$. In Fig. 3 we test
the intermediate values $\lambda=0.8,0.6$ and $0.4$ for both the spectral
function and the positions of the poles (for the latter, $\lambda=0.5$ is also shown). 
\begin{figure}[h]
\begin{center}
\begin{minipage}[b]{7.72cm}
\centering
\includegraphics[width=7.72cm]{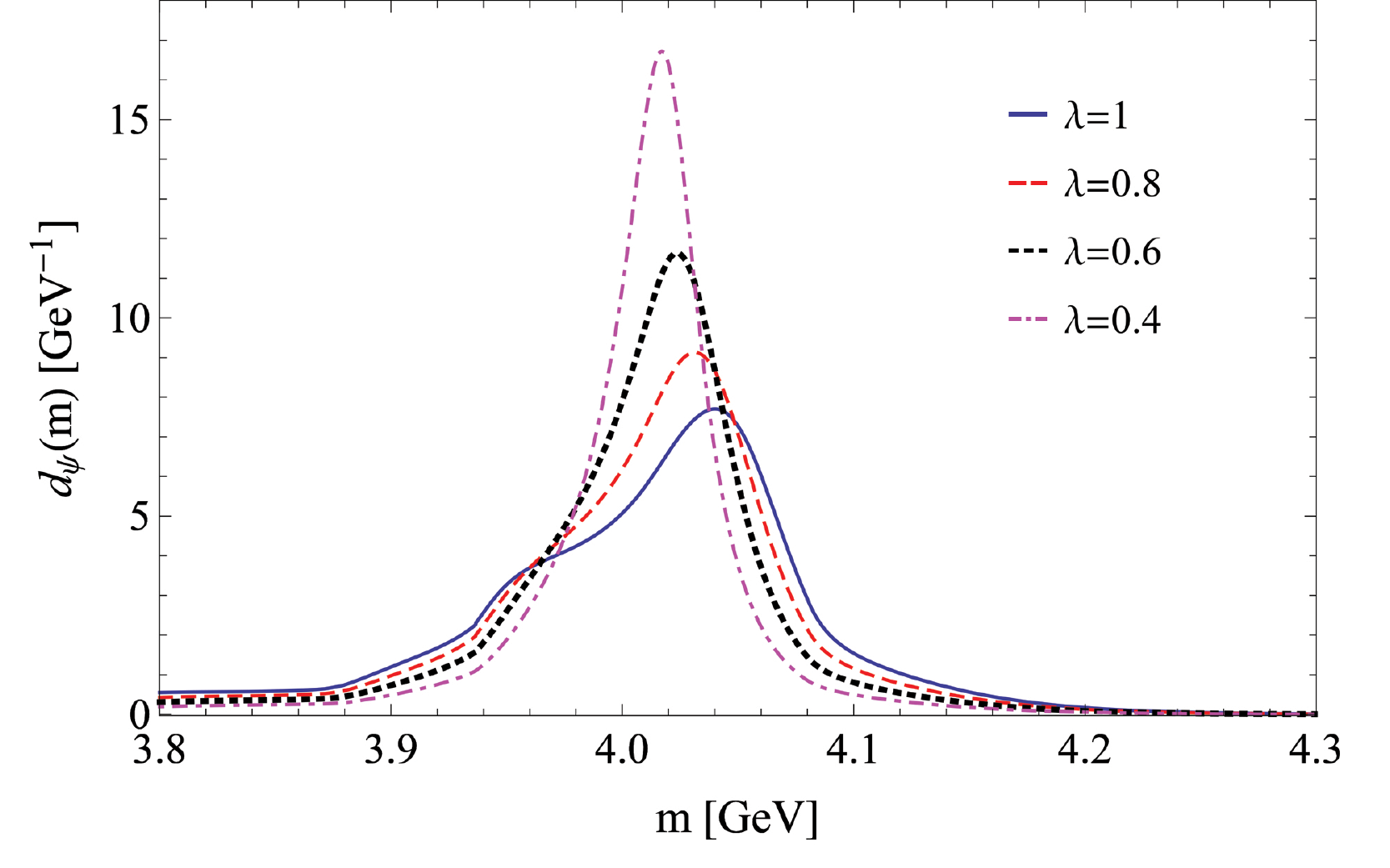}\\\textit{a)}
\end{minipage}
\begin{minipage}[b]{7.72cm}
\centering
\includegraphics[width=7.72cm]{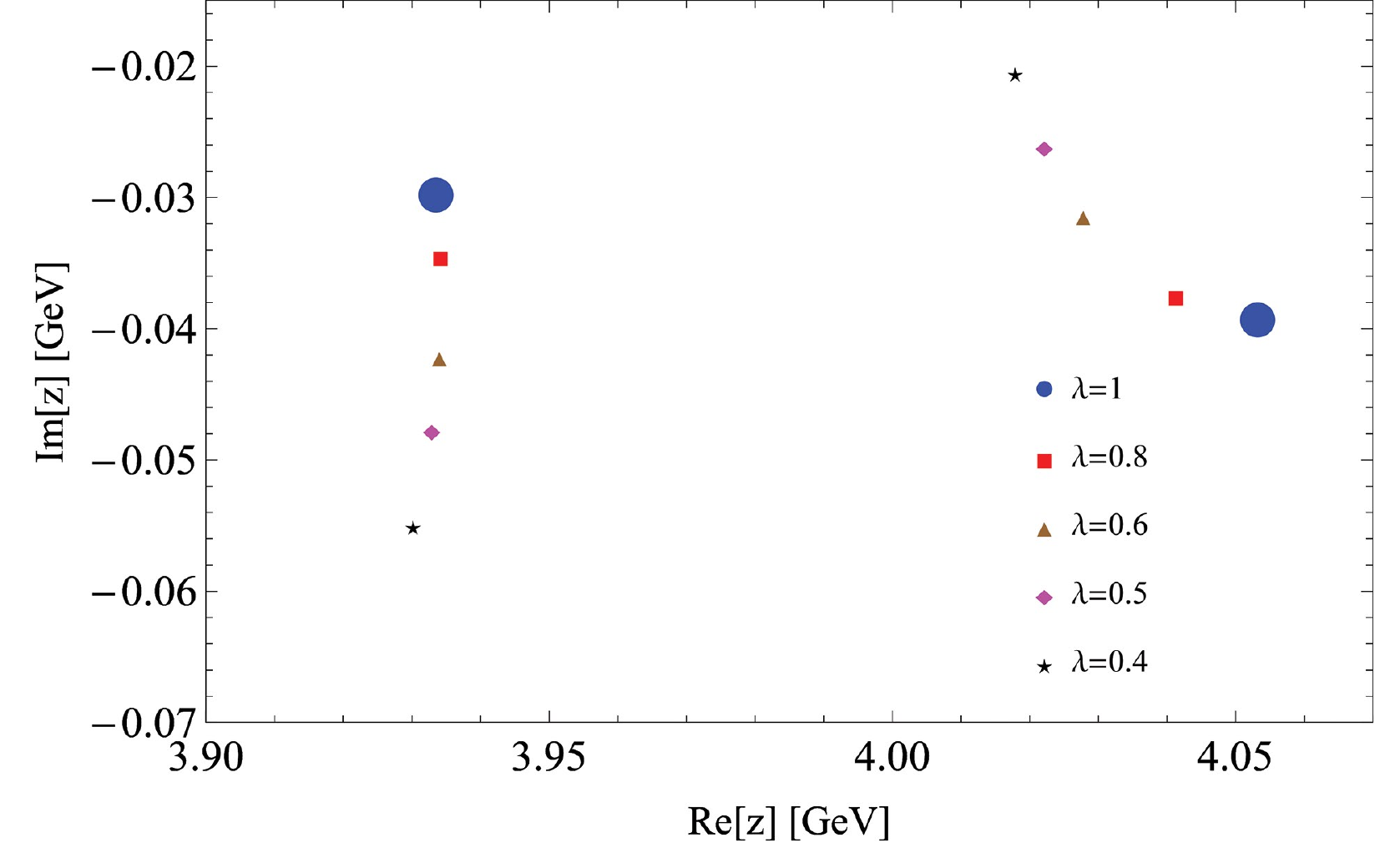}\\\textit{b)}
\end{minipage}
\end{center}
\caption{Study of the changing of the spectral function (panel a) and pole
movement in the complex plane (panel b) for different values of $\lambda$. The
used parameters for the plots are: $g_{\psi DD}=39.84$, $g_{\psi D^{\ast}%
D}=3.44$ GeV$^{-1}$, $g_{\psi D^{\ast}D^{\ast}}=1.85$, $M_{\psi}=4.007$ GeV
and Gaussian form factor $F_{\Lambda}$ with $\Lambda=0.42$ GeV.}%
\end{figure}The large-$N_{c}$ study shows that for smaller $\lambda$ (hence,
larger $N_{c}$), the left enhancement in the spectral function becomes smaller
and finally disappears, while the spectral function becomes narrower. For what
concerns the pole trajectory, the seed pole corresponding to $\psi(4040)$
resonance moves towards to real energy axis, while the additional companion
pole moves away from it. This behavior confirm that the resonance $\psi(4040)$
is a conventional $q\bar{q}$ meson while the second pole is dynamically generated.

As a last point, we comment on mixing with other vector state, in particular
with the closest quarkonium state $\psi(4160).$ By studying the mix propagator (see Refs.
\cite{nonorth,achasov} for some formal equation), we tested how the spectral
function of $\psi(4040)$ changes when taking into account the processes $\psi
(4040)\rightarrow DD^{\ast}$ $\rightarrow\psi(4160)\rightarrow DD^{\ast
}\rightarrow\psi(4040)$ (this is so because $\psi(4160)$ also couples to $DD^*$; analogous processes with \ $DD$ and $D^{\ast}D^{\ast
}$ are possible). The spectral function of $\psi(4040)$ turns out to be only
slightly affected in the region of interest, thus the results here presented
would hold also in the enlarged scenario in which more $\bar{c}c$ states are considered.

\subsection{Decay into $J/\psi\pi^{+}\pi^{-}$}
\label{subsec32}
An important decay channel, in which various $Y$ states have been observed,
among which the $Y(4008)$ state is one, is the $e^{+}e^{-}\rightarrow\gamma J/\psi\pi
^{+}\pi^{-}$ decay, where the photon $\gamma$ comes from ISR. Hence, the reaction
may be recasted into two steps: $e^{+}e^{-}\rightarrow\gamma\left(  e^{+}%
e^{-}\right)  _{\text{off-shell}}$ and $\left(  e^{+}e^{-}\right)
_{\text{off-shell}}\rightarrow J/\psi\pi^{+}\pi.\ $

Since the very same fundamental process is involved, for simplicity we
consider in the following the process $e^{+}e^{-}\rightarrow J/\psi\pi^{+}%
\pi^{-}$ (we thus ignore that the electron-positron pair is off-shell). In
particular, we are interested in the case in which $\psi(4040)$ is an
intermediate state of the reaction
\begin{equation}
e^{+}e^{-}\rightarrow\psi(4040)\rightarrow J/\psi\pi^{+}\pi^{-}\text{ .}%
\end{equation}
There are basically two ways in which this process can take place.\ The first
one involves the emission of two gluons
\begin{equation}
\psi(4040)\equiv c\bar{c}\rightarrow c\bar{c}+gg\rightarrow J/\psi
+f_{0}(500)\rightarrow J/\psi+\pi^{+}\pi^{-}%
\end{equation}
The choice of $f_{0}(500)$ (see \cite{pealaezrev} for a review) as an intermediate
state is motivated by the fact that it is the lightest quantum state with
quantum numbers of the vacuum and is in the right kinematic region
($f_{0}(980)$ is at the border, $f_{0}(1370)$ already too heavy).
Nevertheless, one can repeat the very same discussion by considering different
$f_{0}$ states. This decay mode can be modelled by
\begin{equation}
\mathcal{L}_{\psi jf_{0}}^{direct}=g_{\psi jf_{0}}^{direct}\psi_{\mu}j^{\mu
}f_{0}\text{ ,}%
\end{equation}
where $j^{\mu}$ is the field corresponding to the $J/\psi$ and $f_{0}$ to
$f_{0}(500).$ This term would generate a peak at $4.04$ GeV, which has not been seen
experimentally (in fact, this would be a ``standard'' decay of $\psi(4040)$ peaked at its Breit-Wigner mass). It means that $g_{\psi jf_{0}}^{direct}$ should be quite
small. We will neglect this channel in the following.

There is, however, a second mechanism%

\begin{equation}
\psi(4040)\rightarrow DD^{\ast}\rightarrow J/\psi+f_{0}(500)\rightarrow
J/\psi+\pi^{+}\pi^{-},
\end{equation}
where the additional vertex is represented by the following four-body interaction%
\begin{equation}
\mathcal{L}_{DD^{\ast}jf_{0}}=\lambda_{DD^{\ast}jf_{0}}\left[  \partial^{\mu
}D^{\ast+\nu}D^{-}+\partial^{\mu}D^{\ast0\nu}\bar{D}^{0}\right]  j_{\mu\nu
}f_{0}\text{ ,} \label{lscatf0}%
\end{equation}
where $j_{\mu\nu}=\partial_{\mu}j_{\nu}-\partial_{\nu}j_{\mu}.$
\begin{figure}[h]
\begin{center}
\includegraphics[width=0.55 \textwidth]{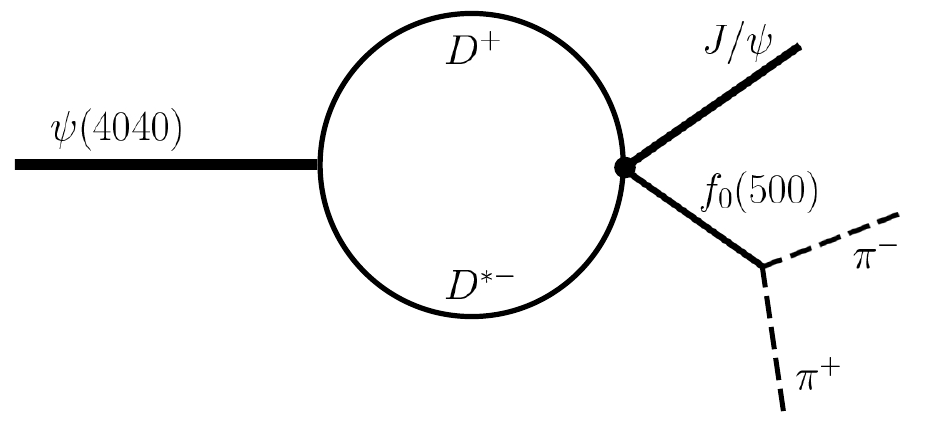}
\end{center}
\caption{Schematic diagram of the decay $\psi(4040)\rightarrow\pi^{+}\pi
^{-}J/\psi$ via $D^{+}D^{\ast-}$ loop.}%
\end{figure}

The spectral function in the channel $J/\psi\pi^{+}\pi^{-}$ reads
\begin{equation}
d_{\psi(4040)\rightarrow J/\psi\pi^{+}\pi^{-}}(m)=\frac{2m}{\pi}\left\vert
\Delta_{\psi}(m^{2})\right\vert ^{2}m\Gamma_{\psi(4040)\rightarrow J/\psi
\pi^{+}\pi^{-}}(m) \label{dpsij}%
\end{equation}
where%
\begin{align}
&  \Gamma_{\psi(4040)\rightarrow J/\psi f_{0}(500)}(m)\nonumber\\
&  =\left\vert g_{\psi jf_{0}}^{direct}+\lambda_{DD^{\ast}jf_{0}}g_{DD^{\ast}%
}\left[  \Sigma_{D^{0}D^{\ast0}}(m^{2})+\Sigma_{D^{+}D^{\ast-}}(m^{2})\right]
\right\vert ^{2}\frac{k}{8\pi m^{2}}\left(  3+\frac{k^{2}}{m_{J/\psi}^{2}%
}\right)  e^{-2\frac{k^{2}}{\Lambda^{2}}}\text{ .}%
\end{align}
with $k\equiv k(m,m_{J/\psi},m_{f_{0}(500)}).$ It is then clear that the
$DD^{\ast}$ loop, together with the Lagrangian $\mathcal{L}_{DD^{\ast}jf_{0}}$
of Eq. (\ref{lscatf0}), generates a mass-dependent coupling for the channel
$\psi(4040)\rightarrow J/\psi f_{0}(500)$.

In general, this decay is small, since both mechanisms are small [they are
suppressed (at least) as $1/N_{c}^{3/2}$ in the large-$N_{c}$ limit], yet the
second mechanism is expected to be dominant in our case. Namely, the seed
state $\psi(4040)$ couples strongly to $D^{\ast}D$ (this is the dominant decay
mode). Moreover, the real part of the $DD^{\ast}$ loops, depicted in Fig. 5.a,
has a pronounced peak at $m_{D}+m_{D^{\ast}}$ at about $3.9$ GeV. The
transition $D^{\ast}D\rightarrow J/\psi f_{0}(500)$ via the Lagrangian
$\mathcal{L}_{DD^{\ast}jf_{0}}$ is rather natural, since it implies a
redistribution of already existing quarks. Moreover, $f_{0}(500)$ couples
strongly to $\bar{u}u$ and $\bar{d}d.$ Hence, in first approximation we shall
neglect $g_{\psi jf_{0}}^{direct}$ in the following.

Summarizing, in Eq. (\ref{dpsij}) the product of two functions is present:
$\left\vert \Delta_{\psi}(m^{2})\right\vert ^{2},$ peaked at $4.04$ GeV, and
$\Gamma_{\psi(4040)\rightarrow J/\psi f_{0}(500)}(m),$ peaked at $3.9$ GeV. Of
course, other channels, such as $DD,$ would also couple $J/\psi f_{0}(500)$,
but (i) the coupling of $\psi(4040)$ to $DD$ is sizably smaller and (ii) the
function $\left\vert \Sigma_{DD}(m^{2})\right\vert ^{2}$ is peaked at the $DD$
threshold, hence the overlap with $\left\vert \Delta_{\psi}(m^{2})\right\vert ^{2}$
is negligible.

\begin{figure}[h]
\begin{center}
\begin{minipage}[b]{7.70cm}
\centering
\includegraphics[width=7.70cm]{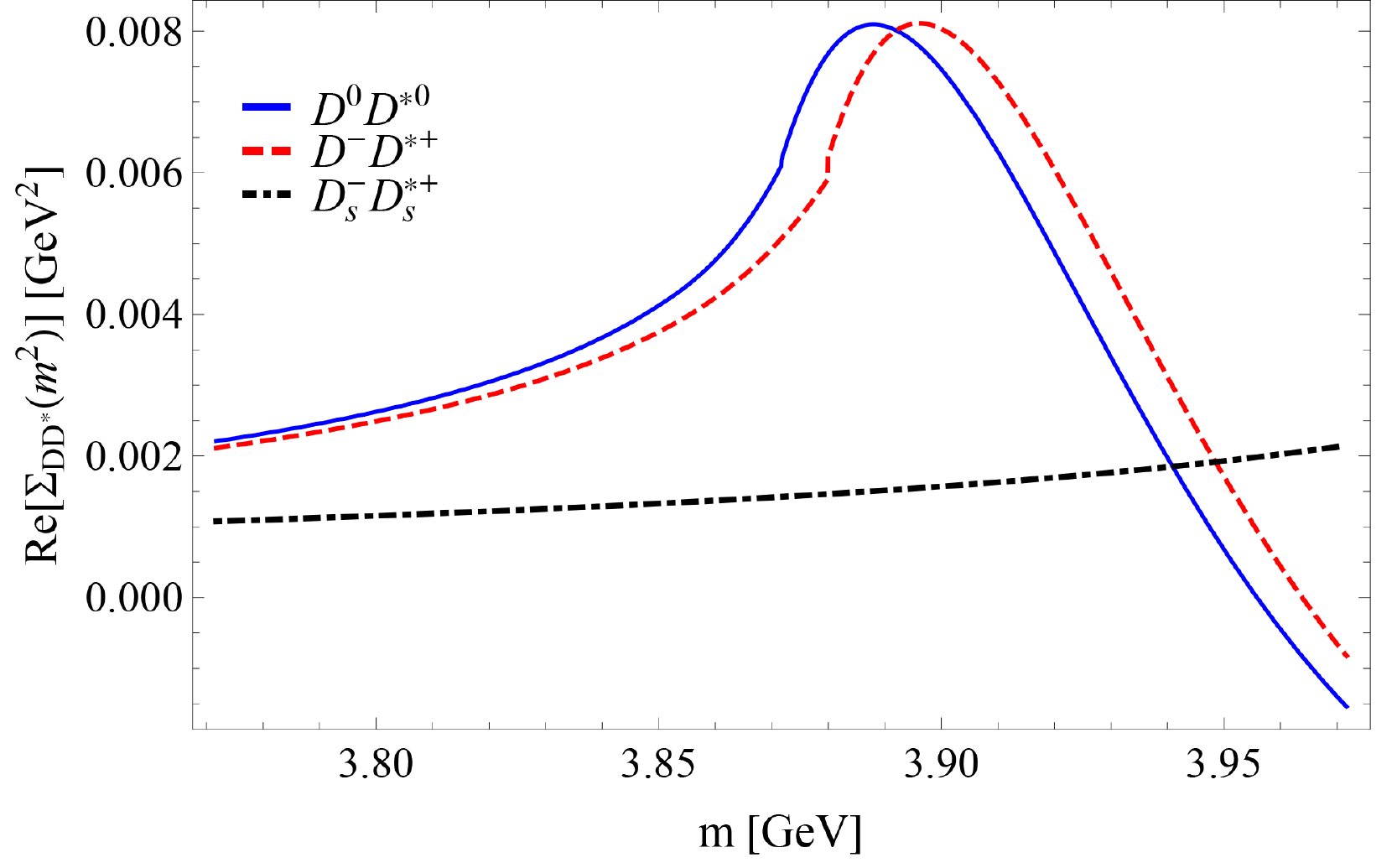}\\\textit{a)}
\end{minipage}
\begin{minipage}[b]{7.70cm}
\centering
\includegraphics[width=7.25cm]{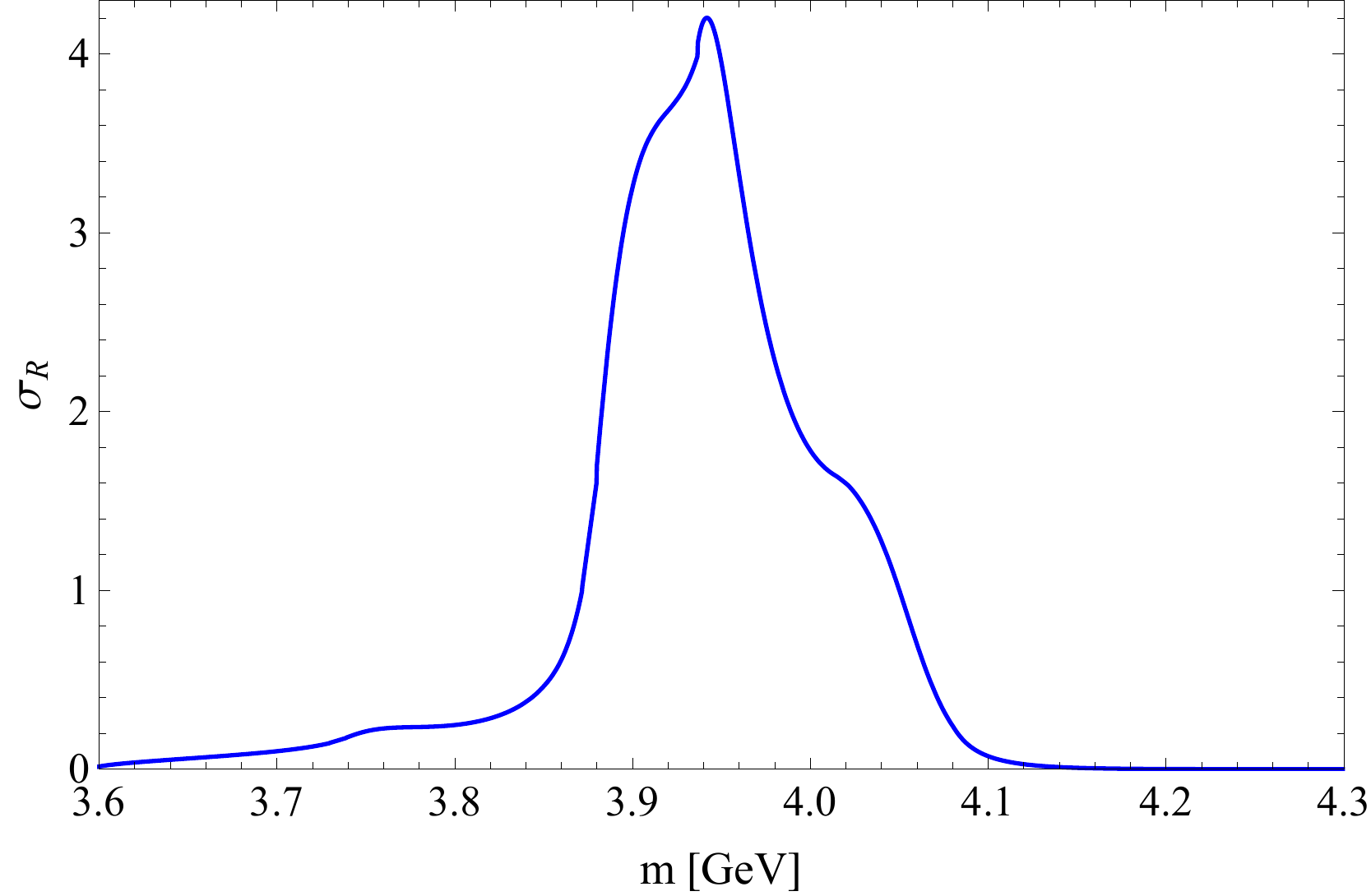}\\\textit{b)}
\end{minipage}
\end{center}
\caption{Left panel: $\operatorname{Re}\Sigma_{D^{0\ast}D^{0\ast}}$ and
$\operatorname{Re}\Sigma_{D^{+\ast}D^{-\ast}}$ as function of $m.$ Both of
them have a peak at about $3.9$ GeV. Right panel: Plot of the (normalized)
cross-section $\sigma_{R}(m)$ of the channel $e^{+}e^{-}\rightarrow
\psi(4040)\rightarrow J/\psi\pi^{+}\pi^{-}$ deifned in\ Eq. (\ref{ratio}). A
deformed and broad structure with a peak at about 3.95 GeV is visible.}%
\label{rys1}%
\end{figure}

For $\sqrt{s}$ in the region of interest, we have
\begin{equation}
\sigma_{e^{+}e^{-}\rightarrow\psi(4040)\rightarrow J/\psi\pi^{+}\pi^{-}%
}\left(  m\right)  =\frac{2\pi}{m}g_{\psi e^{+}e^{-}}^{2}d_{\psi
(4040)\rightarrow J/\psi\pi^{+}\pi^{-}}(m)\text{ .}%
\end{equation}
In Fig. 5.b we plot%
\begin{equation}
\sigma_{R}(m)=\frac{\sigma_{e^{+}e^{-}\rightarrow\psi(4040)\rightarrow
J/\psi\pi^{+}\pi^{-}}\left(  m\right)  }{\sigma_{e^{+}e^{-}\rightarrow
\psi(4040)\rightarrow J/\psi\pi^{+}\pi^{-}}\left(  m=m_{D^{0}}+m_{D^{\ast0}%
}\right)  }. \label{ratio}%
\end{equation}
(In this way, the dependence on the unknown coupling $\lambda_{DD^{\ast}%
jf_{0}}$ cancels and $\sigma_{R}(m_{D^{0}}+m_{D^{\ast0}})=1$). The resulting
form is quite peculiar and is definitely not a simple Breit-Wigner peak. If
the experimental accuracy is not high enough, one may identify this signal as
a broad resonance whose peak is centered at about $4$ GeV. Thus, we suggest
that the `state' $Y(4008)$ is a manifestation of the standard state
$\psi(4040),$ which arises due to the decay into $J/\psi\pi^{+}\pi^{-}$
through the nearby $DD^{\ast}$ loop. It is important to stress that this
conclusion is \textit{independent} of the presence of a dynamically generated
pole and its precise position, but it is a consequence of the strong coupling
to $DD^{\ast}$ and the fact that the \ $DD^{\ast}$ threshold is not far from
the peak of $\psi(4040)$.

This mechanism for the generation of the state $Y(4008)$ does not
necessarily correspond to a resonance. Moreover, the width of the dynamical
pole is not related to the width of the signal of Fig. 5.b. In Appendix B we
also present the results for variations of the parameters, and for a cutoff up
to $0.5$ GeV a very similar outcome is obtained.

\section{Summary and discussion}

We studied the energy region close to the resonance $\psi(4040)$ in the
framework of a QFT effective model. We evaluated its spectral function and
found that, besides the expected resonance pole of $\psi(4040)$ (corresponding
to peak in the spectral function and to the underlying seed $c\bar{c}$ state),
an additional companion pole (no peak, but an enhancement in the spectral
function) emerges naturally within our approach (see Table 1). Illustrative
result in agreement with phenomenology are: $4.053-i0.039$ GeV for the seed
state $\psi(4040)$ and $3.934-i0.030$ GeV for the enhancement. A large-$N_{c}$
study confirms that $\psi(4040)$ resonance is predominantly a charm-anticharm
state, while the second pole is dynamically generated by meson-meson quantum fluctuations.

The pole itself cannot be directly associated to the $Y(4008).\ $Namely, this
pole would be mostly visible in a two-peak structure of the cross-section
$e^{+}e^{-}\rightarrow\psi(4040)\rightarrow DD^{\ast}$ (whose data precision
is not good enough). Yet, the chain $e^{+}e^{-}\rightarrow\psi
(4040)\rightarrow DD^{\ast}\rightarrow J/\psi+f_{0}(500)\rightarrow J/\psi
\pi^{+}\pi^{-}$is quite promising: the $DD^{\ast}$ loop generates a peak at
about $3.9$ GeV in the cross-section. The strong coupling of $\psi(4040)$ to
$DD^{\ast}$ and the overlap of the $DD^{\ast}$-loop function with the modulus
square of the propagator are responsible for a quite broad peak in the
corresponding spectral function, which can be identified with $Y(4008)$. The
important point is that the existence of an additional pole corresponding to
$Y(4008)$ is possible (and indeed it does exist for the parameters of our
model) but is actually not necessary for the process that we describe. The
very same mechanism can be also investigated in the future in other channels,
as for instance in connection with the states $Y(4260)$ and $\psi(4160)$.

\bigskip

\textbf{Acknowledgements} The authors thank S. Coito for useful discussions.
M.P. and F.G. acknowledge financial support from the Polish National Science
Centre (NCN) through the OPUS project no. 2015/17/B/ST2/01625. P. K. were supported by the Hungarian OTKA fund K109462 and by the ExtreMe Matter Institute EMMI at the GSI Helmholtzzentrum fur Schwerionenforschung, Darmstadt, Germany.

\appendix

\section{Decay widths on shell}

Here we present the results for the on-shell decay widths for both form
factors and for different values of the cutoff, respectively. Even if the
qualitative picture does not change much, one observes non-negligible
variations of the partial decay widths as function of the cutoff $\Lambda.$ In
the future, a better determination of the decay $\psi(4040)$ into $D_{s}%
^{+}D_{s}^{-}$ (presently only seen) would constitute a useful constraint on
our model.

\begin{table}[h]
\centering
\par
\renewcommand{\arraystretch}{1.25}
\begin{tabular}
[c]{|c|c|c|c|}\cline{3-4}%
\multicolumn{2}{c|}{} & \multicolumn{2}{c|}{\textbf{Partial decay width
[MeV]}}\\\hline
$\mathbf{\Lambda}$ \textbf{[GeV]} & \textbf{Decay channel} & \textbf{Gaussian
form factor} & \textbf{Dipolar form factor}\\\hline
& $DD$ & $4.24 \pm0.80$ & $8.8\pm3.5$\\
& $D_{s}D_{s}$ & $55 \pm10$ & $28 \pm11$\\
0.4 & $D^{*}D$ & $17.7 \pm9.3$ & $37 \pm12$\\
& $D^{*}_{s}D_{s}$ & $0$ & $0$\\
& $D^{*}D^{*}$ & $3.2_{-3.2}^{+3.3}$ & $6.6 \pm6.1$\\\hline
& $DD$ & $5.6 \pm1.4$ & $9.2 \pm3.8$\\
& $D_{s}D_{s}$ & $47 \pm12$ & $26 \pm10$\\
0.42 & $D^{*}D$ & $23 \pm11$ & $38 \pm12$\\
& $D^{*}_{s}D_{s}$ & $0$ & $0$\\
& $D^{*}D^{*}$ & $4.2^{+4.3}_{-4.2}$ & $6.9 \pm6.3$\\\hline
& $DD$ & $7.5 \pm2.5$ & $9.8 \pm4.2$\\
& $D_{s}D_{s}$ & $35 \pm12$ & $22.2 \pm9.6$\\
0.45 & $D^{*}D$ & $31 \pm12$ & $41 \pm12$\\
& $D^{*}_{s}D_{s}$ & $0$ & $0$\\
& $D^{*}D^{*}$ & $5.7 \pm5.4 $ & $7.3 \pm6.6$\\\hline
& $DD$ & $9.8 \pm4.2$ & $10.7 \pm5.0$\\
& $D_{s}D_{s}$ & $22.2 \pm9.6$ & $17.0 \pm8.0$\\
0.5 & $D^{*}D$ & $41 \pm12$ & $44 \pm12$\\
& $D^{*}_{s}D_{s}$ & $0$ & $0$\\
& $D^{*}D^{*}$ & $7.3 \pm6.6$ & $8.0 \pm7.1$\\\hline
& $DD$ & $11.8 \pm6.2$ & $11.9 \pm6.3$\\
& $D_{s}D_{s}$ & $10.4 \pm5.4$ & $9.6 \pm5.1$\\
0.6 & $D^{*}D$ & $49 \pm11$ & $50 \pm11$\\
& $D^{*}_{s}D_{s}$ & $0$ & $0$\\
& $D^{*}D^{*}$ & $8.8 \pm7.6$ & $8.9 \pm7.7$\\\hline
\end{tabular}
\caption{Partial decay widths on-shell for both types of form factor and
different values of $\Lambda$ parameter.}%
\end{table}

\section{Decay into $J/\psi\pi^{+}\pi^{-}$ - variations of the parameter $\Lambda$}

\begin{figure}[h]
\begin{center}
\includegraphics[width=0.65 \textwidth]{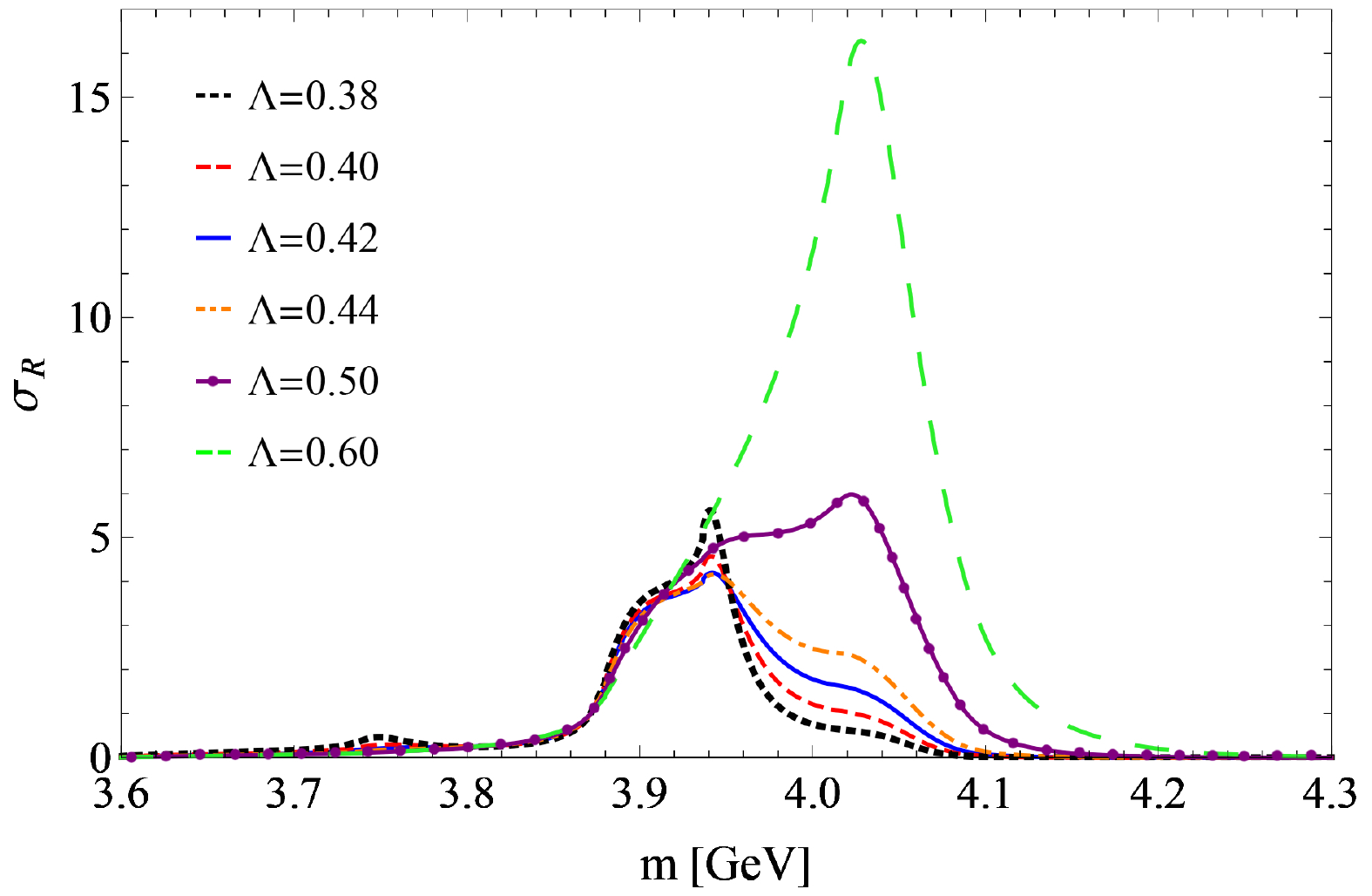}
\end{center}
\caption{Study of the cross-section ratio of Eq. (\ref{ratio}) upon changing the
value of cutoff parameter.}%
\label{crosssection}%
\end{figure}

As it was discussed in the paper, the decay of $\psi(4040)$ into $J/\psi
\pi^{+}\pi^{-}$ via $DD^{\ast}$ loops generates a sizable peak in the
cross-section in the energy region close to $4$ GeV. This is an important
aspect of our theoretical framework, thus we present in\ Fig.
\ref{crosssection} how the results of the cross-section of Eq. (\ref{ratio})
depend on the different values of cutoff. 

When $\Lambda$ varies in the range from $0.38$ up to (at most) $0.5$ GeV one
observes a broad peak at about $3.95$ GeV. For $\Lambda=0.5$ one has actually a
quite broad structure, but the peak is already at bout 4.04 GeV. For larger
values of $\Lambda$, there is a single peak close to $4.04$ GeV which
corresponds to standard $c\bar{c}$ seed state $\psi(4040)$. Hence, for the
generation of a signal resembling $Y(4008)$ the value of $\Lambda$ should not
exceed $0.5$ GeV. \newline

\end{document}